# A Virtual Java Simulation Lab for Computer Science Students

Díaz, Javier - Queiruga, Claudia - Claudia Villar - Fava, Laura
jdiaz, claudiaq, lfava@info.unlp.edu.ar - cvillar@netverk.com.ar
CONICET, LINTI – Universidad Nacional de La Plata - Calle 50 y 115 – 1º Piso. La Plata (1900) Argentina

**Abstract:** The VJ-Lab is a project oriented to improve the students learning process of Computer Science degree at the National University of La Plata. The VJ-Lab is a Web application with Java based simulations. Java can be used to provide simulation environments with simple pictorial interfaces that can help students to understand the subject. There are many fields in which it is difficult to give students a *feel* for the subject that they are learning. Computer based simulations offer a fun and effective way to enable students to **learn by doing**. Both, practicing skills and applying knowledge are both allowed in simulated worlds. We will focus on the VJ-Lab project overview, the work in progress and some Java based simulations running. They imitate the behavior of data network protocol and data structure algorithms. These applets are produced by the students of the 'Software Development Laboratory' [i] course.

## Introduction

Since 1998 the ´Software Development Laboratory´ [ii] course has focused on newer JAVA aspects. As a final project of the course, students developed educational applets for the purpose of illustrating data structure algorithms and data network protocols.
The applets include highly configurable interfaces, animations and a high degree of interaction in order to attract user´s attention. The result was quite interesting in terms of the generated codes, applets observable features and the students meaningful learning. Then, we considered to use the developed applets as educational material in ´Data Network Seminar´ [iii], ´Data Network Technology` [iv] and `Data Structure` courses. From this experience we decide to build the VJ-Lab in order to promote knowledge applications of Java with Computer Science curricular contents.

## The VJ-Lab Project

*Goals and Learning Issues*

The VJ-Lab was conceived to promote undergraduate students of Computer Science to: a.-accessing to the newest technologies in an easy and fast way; b.-training in Java technology from a meaningful viewpoint; c.-facilitating and automating the complexity of contents and processes (e.g. by computer based simulation of IP packages routing in a WAN by a single PC); d-handling a vast amounts of complex data by Java simulations (e.g. Hierachical Tree AVL, Topological Sort, etc.); e.-learning by doing computer science curricular contents. The two approaches in this learning process are: a) to apply high order abilities by using processes simulations, which are more difficult to understand from traditional teaching and learning tools; b) to putt into practice the Java concepts taught and apply theoretical curricular contents by building Java-simulations.
*Teachers must teach students to do things, rather than having them be told about what others have done* [v].
*Learning* is the accumulation and indexing of cases. *Thinking* is to find and consider an old case to apply in the decision-making of a new case or situation. To make thinking beings, we must encourage explanation, exploration, generalization, and case accumulation. Learning is essentially a discovery process, a natural act and we are all natural learners. To make learning by doing possible it is necessary to transform educational activities so that it looks, feels, and is like doing. The use of *computer based simulations* have potential values and benefits to facilitate the learning processes of the students (Cairns, K. 1995). It creates a realistic computer environment that makes *learn by doing possible*: a.-providing opportunities for active experimentation in solving realistic problems which require the integration of knowledge, skills, personal attitudes, and positive work values; b.-formulating and testing hypotheses, identify patterns in their own and others' behavior, making decisions and observing consequences; c.- to observe the impact of such changes, as well as, modifying decisions and actions; d.-correcting errors, positive feedback. Simulations elicit higher levels of arousal, motivation, task engagement, and quality of problem-solving in students more than traditional classroom methods. By simulations teachers can review student's progress and focus on their achievments and/or weaknessess [vi].

Contents

The topics covered in the VJ-Lab comes from: a.-Data Networks Course: RIP (Routing Information Protocol), Traceroute program operation, X.25 protocol (link and network layers), ARP (Address Resolution Protocol), RARP (Reverse Address Resolution Protocol) and Ping program (Packet InterNet Groper); b.-Data Structure Course: Binary Search Tree AVL, Topologhycal Sort, MinHeap, MaxHeap, Dijkstra algorithm.

*Technical Features*

The VJ-Lab is a web-based system centered on JAVA$^{TM}$ specifically focusing in developing applets in Java 2. It introduces the first deployment of the most significant cross-platform GUI technology.
The Java Foundation Classes are made up of several technologies: AWT (Abstract Windowing Toolkit), Swing, Accesibility and Java 2D. These technologies are the core of Java's UI support. The JFC contains a powerful, mature delegation event model, printing clipboard support, a lightweight UI framework and it is 100% JavaBeans [vii] compliant. Swing Components: Swing extends the original AWT by adding a comprehensive set of graphical interfaces class libraries that is completely portable and it is delivered as part of the Java Platform. The Swing components improve the GUI development providing high quality GUI components which are peerless or lightweight, look and feel pluggable, customizable and transparent. Additionally, the performance was greatly improved since the incorporation of lightweight components and the new event management model.

## Work in Progress

The training applets depicted in Table 1 are visual, some of them have iconic aspects and others make use of direct manipulation. They are highly configurable, that is, the users set specific data protocol, sound, images, colors, etc. They are developed using Swing components, Java 2D, some of them support multithread programming and Java Data Base Connection. The interaction techniques used in the applets are the following: a) in ARP, RIP and one of the Traceroute version applets the final user interacts with a given network without the need of introducing any setting data, b) in the other version of Traceroute, RIP and X.25 applets, the final user is obliged to build and configure the data network. In this case, the user must define the IP addressing, the links to join the equipments, the tables used for routing algorithms, etc. When all of these datas are ready, the applets run the specific protocol. The criteria used by professors to evaluate the applet GUI were organization, economy and communication, according to Nielsen principles (http://www.useit.com/papers/heuristic/) and the Schneiderman [viii] guidelines. In applets mentioned it is possible to take network actions such as, interrupting *the animation* protocol and then resuming the animation in the same status, *introduce noises* in the data package, *break a link, put on/off* a host, *change the parameters* of the protocol. In addition, it is possible to take actions that enable to change the user's interface. The final user sets the applet aparience such as models of routers and hosts, foregrounds and backgrounds colors.

The appplets implemented by the students in Software Development Laboratory course were produced according to the following steps: a.-students work in teams; b.-professors write the final projects; c.-students have three opportunities to present the applets; d.-professors correct them from two points of view: functionality and graphics user interface aspects; e.-the applets which reach the expectatives goals are selected and published in the university website.

| | |
|---|---|
| Data Network | http://www.linti.unlp.edu.ar/catedras/Laboratorio/1999/applets.htm |
| Data Structure | http://www.linti.unlp.edu.ar/catedras/Laboratorio/1998/applets.htm |
| RIP* | http://www.linti.unlp.edu.ar/catedras/Laboratorio/1999/finales/Russo-Cibran/Applet.htm |
| | http://www.linti.unlp.edu.ar/catedras/Laboratorio/1999/finales/Lastra-Maceira/Applet.htm |
| Traceroute* | http://www.linti.unlp.edu.ar/catedras/Laboratorio/1999/finales/Bazzoco/Applet.htm |
| X.25* | http://www.linti.unlp.edu.ar/catedras/Laboratorio/1999/finales/Silva-Macia/Applet.htm |
| ARP* | http://www.linti.unlp.edu.ar/catedras/Laboratorio/1999/finales/Rosso/Applet.htm |
| | http://www.linti.unlp.edu.ar/catedras/Laboratorio/1999/finales/JDiaz/Applet.htm |
| RARP* | ttp://www.linti.unlp.edu.ar/catedras/Laboratorio/1999/finales/Martinez-Feito/Applet.htm |
| PING* | http://www.linti.unlp.edu.ar/catedras/Laboratorio/1999/finales/Mastricardi/Applet.htm |

**Table 1. Applets simulations** *Applets specifc topics*

## Up coming developments

The short term develop actions are aimed at increasing the amount of Java-simulation in data network and data structure subjects according to curricula. Other activities will be: a.-making empirical studies to measure the effectiveness of the V-J Lab in order to improve learning skills in the area of Computer Science; b.-implementing usability studies to asses the effectivity of the applets; c.-producing standards guides for future user interface designs; d.-building a front – end system based on the Web for facilitating and supporting the students use.